\documentstyle[12pt]{article}
\begin{document}
\vskip 2cm
\begin{center}
{\sf {\Large Novel symmetries in the modified version of two dimensional Proca theory }}

\vskip 3.0cm

{\sf T. Bhanja$^{(a)}$, D. Shukla$^{(a)}$, R. P. Malik$^{(a,b)}$}\\
$^{(a)}$ {\it Physics Department, Centre of Advanced Studies,}\\
{\it Banaras Hindu University, Varanasi - 221 005, (U.P.), India}\\

\vskip 0.1cm

{\bf and}\\

\vskip 0.1cm

$^{(b)}$ {\it DST Centre for Interdisciplinary Mathematical Sciences,}\\
{\it Faculty of Science, Banaras Hindu University, Varanasi - 221 005, India}\\
{\small {\sf {e-mails: tapobroto.bhanja@gmail.com; dheerajkumarshukla@gmail.com;  rpmalik1995@gmail.com}}}

\end{center}

\vskip 2cm

\noindent
{\bf Abstract:} By exploiting the Stueckelberg approach, we obtain a gauge theory for the two (1+1)-dimensional (2D) Proca theory and demonstrate that this theory is endowed with, in addition to the 
$usual$ Becchi-Rouet-Stora-Tyutin (BRST) and anti-BRST symmetries, the 
on-shell nilpotent (anti-)co-BRST symmetries, under which, the {\it total} gauge-fixing term remains invariant. The anticommutator of the BRST and co-BRST  (as well as anti-BRST and anti-co-BRST) symmetries define a {\it unique} bosonic symmetry in the theory, under which, the ghost part of the Lagrangian density remains invariant. To establish connections of the above symmetries with the Hodge theory, we invoke a pseudo-scalar field in the theory. Ultimately, we demonstrate that the full theory provides a field theoretic example for the Hodge theory where the continuous symmetry transformations provide a physical realization of the de Rham cohomological operators and discrete symmetries of the theory lead to the physical realization of the Hodge duality operation of differential geometry. We also mention the physical implications and utility of our present investigation.
\vskip 0.5cm
\noindent

\vskip 0.2cm
\noindent

\newpage

\noindent
\section {Introduction}

The well-known Proca field  theory is a generalization of the Maxwell field theory where a
vector boson has three degrees of freedom\footnote{This statement is true only
in the physical four dimensions of spacetime.} 
due to the presence of its mass. The latter attribute 
of such a vector boson spoils
the beautiful gauge symmetry transformations of Maxwell's theory. In modern language, any arbitrary
(e.g. Maxwell) gauge theory is endowed with the first-class constraints whereas the Proca theory is characterized by the
second-class constraints in the terminology of Dirac's prescription for the classification
scheme of constraints (see, e.g. [1,2] for details). As a consequence, Proca theory does not respect the
gauge symmetry invariance (because of the presence of the mass).
However, Stueckelberg's formalism
restores the gauge invariance in the massive gauge (i.e. Proca) theory where the mass and gauge invariance 
co-exist together
because of the presence of Stueckelberg's field in addition to the spin-1 vector field of the theory
(see, e.g. [3]).

The purpose of our present investigation is to take the specific case of two (1 + 1)-dimensional (2D)
modified version of the Proca field theory (that incorporates the Stueckelberg field) and study its various
symmetry properties within the framework of Becchi-Rouet-Stora-Tyutin (BRST) formalism. We demonstrate
that a set of {\it six} continuous symmetries and a couple of mathematically {\it useful} discrete
symmetries exist for the modified version of the massive Proca gauge field theory. In fact, we show that,
in addition to the usual (anti-)BRST symmetries, the theory respects (anti-)co-BRST symmetries, a 
unique bosonic symmetry, the ghost-scale symmetry and a couple of discrete symmetries. As a consequence
of the above symmetries, the full Lagrangian density of the theory is {\it uniquely}\footnote{In fact, 
we have two equivalent Lagrangian densities [cf. (10),(11) below] for our present theory. The key feature
of these Lagrangian densities is that the signature of all the terms is fixed due to the 
presence of many symmetries in the theory. 
In this respect, these Lagrangian densities are {\it unique}.} defined with definite
signatures associated with all the terms and it becomes a model for the Hodge theory where the continuous set
of symmetries provide a physical realization of the {\it abstract} de Rham 
cohomological operators\footnote{ On a compact manifold without a boundary, there exists a set 
of three cohomological operators ($d, \delta, \Delta$) where $ d = dx^\mu \partial_\mu$ and $\delta = \pm * d *$
are the nilpotent ($d^2 = \delta^2 = 0$)
exterior and co-exterior derivatives and $\Delta = (d + \delta)^2 = d \delta + \delta d$ is the
Laplacian operators [4-10]. Here $(*)$ is the Hodge duality operation.}  and discrete
symmetries give physical meanings to the {\it abstract} Hodge duality operation of differential geometry.

The nilpotent (anti-)BRST symmetry transformations for the gauge invariant version of Proca theory exist
in any arbitrary dimension of spacetime (cf. Sec. 2). One of the decisive features of these symmetries is
the observation that the total kinetic term, owing its origin to the exterior derivative, remains invariant.
On the other hand, as we have demonstrated in our present endeavor, the nilpotent (anti-)co-BRST symmetries exist
in the two (1 + 1)-dimensions of spacetime for the Abelian 1-form massive gauge theory, under which the total gauge-fixing term
(owing its origin to the co-exterior derivative) remains invariant. There exists a {\it unique} bosonic 
symmetry in the 2D theory, under which the Faddeev-Popov ghost terms remain invariant. In our present 
investigation, we have derived an extended version of the BRST algebra and have shown that this algebraic structure
is exactly like that of the de Rham cohomological operators of differential geometry.

We have also shown, in our present endeavor, that there is two-to-one mapping between the 
generators of the continuous symmetry
transformations of the theory and cohomological operators. A couple of discrete symmetries of our present theory
have been shown to be the analogue of the Hodge duality operation of differential geometry. In principle, we can
have many discrete symmetry transformations in the theory. However, we have concentrated only on two discrete symmetry
transformations, in our present endeavor, which are {\it useful} to us in providing the analogue of the
relationship $\delta = \pm * \, d\, *$ in the language of the interplay between the continuous and
discrete symmetry transformations. We have also shown that the role of the degree of a differential form
is played by the ghost number of a state in the quantum Hilbert space of states of our present theory.

The main motivating factors behind our present investigation are as follows. First, our present 2D model is the one
where {\it mass} of the gauge field and (anti-)BRST symmetries, (anti-)co-BRST symmetries, a bosonic symmetry, the
ghost-scale symmetry and a set of discrete symmetries co-exit {\it together}. Second, the modified version of Proca
theory is radically {\it different} from the models of gauge theories [11-21] and ${\cal N} = 2$ supersymmetric
quantum mechanical models [22-24] which have been shown to be the examples of Hodge theory. Finally, our present
endeavor is our modest step towards our main goal of finding $massive$ models for the Hodge theory in physical
four (3 + 1)-dimensions of spacetime  where, perhaps, 1-form gauge field and higher $p$-form ($ p = 2, 3, 4,...)$ gauge fields will merge together in a meaningful manner to produce the (anti-)BRST 
invariant {\it massive} gauge theories.

In broader perspective, besides the above motivations, our present 
kind of studies have been physically meaningful because,
exploiting the ideas of (anti-)BRST, (anti-)co-BRST and bosonic symmetries, we have been able to prove that the 2D
free (non-)Abelian theories (without any interaction with matter fields) [25-28], are a new type of topological field
theories (TFTs) which capture some aspects of Witten type TFT and a part of Schwarz type TFT. Furthermore,
utilizing the key properties of Hodge theory (especially its symmetries and conserved charges), we have been able to 
demonstrate that the 4D free Abelian 2-form and 6D Abelian 3-form gauge theories [19-21] are quasi-TFTs. Thus, our
present endeavor encompasses in its folds mathematically as well as physically interesting results.

 The contents of our investigation are organized as follows. In Sec. 2, we concisely recapitulate the bare essentials of the Proca theory, its generalization to a gauge theory by Stueckelberg's formalism and its on-shell nilpotent (anti-)BRST symmetries in any arbitrary dimension of spacetime. Our Sec. 3 is devoted to the discussion of the existence of on-shell nilpotent (anti-)co-BRST symmetries in two (1+1)-dimensions of spacetime. We elaborate on the derivation of a {\it unique} bosonic symmetry for this 2D theory from the anticommutators of (anti-)BRST and (anti-)co-BRST symmetries in Sec. 4. We discuss the discrete as well as ghost-scale symmetries of our present theory in Sec. 5. In Sec. 6, we establish connections between the symmetry operators (as well as  their corresponding conserved charges) and the cohomological operators of differential geometry. Finally, we make some concluding remarks 
and point out a few future directions in Sec. 7.

In our Appendices A and B, we discuss about a couple of non-nilpotent supersymmetric type symmetry transformations and a unique bosonic symmetry
transformation for the (anti-)BRST invariant Lagrangian density (8) (see below). Our Appendix C is devoted to to the discussion of a few symmetries
of the Lagrangian density ${\cal L}_{(b_2)}$ (cf. (11)).

{\it General conventions and notations:} Throughout the whole body of our text, we shall use the notations
for the (anti-)BRST and (anti-)co-BRST symmetry transformations as $s_{(a)b}$ and $s_{(a)d}$, respectively. Similarly, 
we shall also adopt the notations for the ghost-scale and bosonic symmetry transformations as $s_g$ and $s_\omega$,
respectively. These notations would be used for the 
two {\it equivalent} Lagrangian densities that are present in our theory.
Other notations would be clarified at appropriate places and we shall focus only on the {\it internal} symmetries
and shall {\it not} even touch anything connected with the {\it spacetime} symmetries. We shall assume
that the spacetime background manifold is {\it flat} and {\it Minkowskian}.

\section {Preliminary: Proca theory as a gauge theory and on-shell nilpotent (anti-)BRST symmetries}

Let us begin with the following Proca Lagrangian density ${\cal L}_0 $ for a massive 
boson (with mass $\it m$) in any arbitrary dimension of spacetime (see, e.g. [3] for details)
\begin{eqnarray}
{\cal L}_0 = -\frac {1} {4} F^ {\mu \nu} \,F_ {\mu \nu} + \frac {m^2} {2}\, A_\mu\, A^ \mu,
\end{eqnarray} 
where the curvature tensor $F_ {\mu \nu} = \partial _\mu A_\nu - \partial _\nu A_ \mu $ has been derived from the 2-form 
$F^{(2)} = d A^{(1)} = [(dx^\mu \wedge dx^\nu)/2!]\, F_{\mu\nu} $ where the 1-form $ A^{(1)} = dx^{\mu}\, A_ \mu $ defines the bosonic field 
$A_\mu$. Here $d= dx^\mu \partial_\mu$ (with $ d^{2} = 0$) 
is the exterior derivative and Greek indices $\mu , \nu , \lambda ... = 0, 1, 2, 3..., (D-1) $ in $D$-dimensions of spacetime. 
It can be shown that the physical system (1) is endowed with the second-class constraints in the language of Dirac's prescription for 
the classification of constraints [1,2]. As a consequence, there is {\it no} gauge symmetry in the theory 
as this symmetry is generated  $only$ by the first-class constraints. The existence of the latter is a key signature of a gauge theory [1,2].
Exploiting the celebrated Stueckelberg's approach [3], we can replace: 
$A_\mu\rightarrow A_\mu - ({1/m})\,{\partial_\mu \phi}$ 
which results in a kinetic term for the real scalar field ${\phi}$. 
Thus, the above Lagrangian density (${\cal L}_0$)  takes another form ($ {\cal L}_s $) as: 
\begin{eqnarray}
{\cal L}_s = - \frac {1}{4}\, F^{\mu\nu} \,F_ {\mu \nu} + \frac {m^2}{2} \,A_\mu \,A^\mu 
+ \frac {1}{2} \,\partial_\mu \phi\,\partial^\mu \,\phi - m \,A_\mu \,\partial^\mu \,\phi \,.
\end{eqnarray}
The above Lagrangian density (${\cal L}_s $) respects the following infinitesimal, local and 
continuous gauge symmetry transformations $({\delta_ g})$, namely;
\begin{eqnarray} 
\delta_g\, A_\mu = \partial_ \mu\, \Lambda ,\qquad \qquad \delta_g\, \phi = m\, \Lambda , 
\end{eqnarray}
because     $ \delta_g \,{\cal L}_s=0 $ .
The parameter ${\Lambda}$ in (3) is a local gauge parameter. We state, in passing, that the second-class constraints of 
the original Proca theory (cf. (1)) have already been converted into the first-class constraints because of the presence 
of the ordinary scalar field $\phi$, in addition to the 
bosonic  field $A_\mu$, in our modified Lagrangian density (2).

The above ``classical" local, continuous and infinitesimal gauge symmetry transformations (3) can be generalized to the ``quantum" level. The latter are the on-shell nilpotent $(s_{(a)b}^2 = 0 )$ (anti-)BRST symmetry transformations $s_{(a)b}$ :
\begin{eqnarray}
&& s_{ab} \,A_\mu = \partial_\mu\, \bar C,\qquad  \,\,s_{ab}\, \bar C =0,\qquad \quad s_{ab}\, C = i\,(\partial\cdot A + m \phi ),\nonumber\\
&& s_{ab}\,\phi = m \,\bar C,\qquad\quad s_{ab}\,F_{\mu\nu} = 0,
\qquad \,\,s_{ab}\,(\partial\cdot A \,+ m \,\phi)= (\Box + m^2)\,\bar C,\nonumber\\ 
&& s_{b}\, A_\mu = \partial_\mu \,C, \,\qquad \,\,s_b \,C = 0, \qquad 
\quad \, \,\, s_{b}\,\bar C = -i\,(\partial\cdot A + m\,\phi ), \nonumber\\
&&s_{b}\, \phi = m\,C, \qquad \quad \,s_{b}\, F_{\mu\nu} = 0, \qquad \quad s_{b}\, 
(\partial\cdot A + m\, \phi) = (\Box + m^2)\, C ,
\end{eqnarray}
for the ``quantum'' generalized version (${\cal L}_B$) of 
the Lagrangian density (${\cal L}_s$) which incorporates the gauge-fixing as  well as 
Faddeev-Popov ghost terms as given below:
\begin{eqnarray}
{\cal L}_B &=& -\frac {1} {4}\, F^ {\mu \nu}\, F_ {\mu \nu} + \frac{m^2} {2}\, A_\mu\, A^ \mu + 
\frac {1}{2}\, \partial_\mu \,\phi\, \partial^\mu \,\phi - m\, A_\mu\, \partial^\mu\, \phi \nonumber\\ 
&-& \frac {1}{2}\,(\partial\cdot A + m\,\phi)^2 - i \,\partial_\mu\,\bar C\,\partial^\mu\, C+ i\, m^2\, \bar C\, C,
\end{eqnarray}
where we have defined $ ( \partial\cdot A ) = \partial_\mu A^\mu $ and the fermionic ($C^2=\bar C^2=0,C \bar C+ \bar C C=0 $) (anti-)ghost field (${\bar C}$)$C$ are needed for the validity of unitarity in the theory.

The gauge fixing term $ [-(\partial\cdot A + m\phi)^2/2] $ 
has its origin in the co-exterior derivative $\delta = \pm * d *$ \,
(with $ \delta^2 = 0 $) where  the Hodge duality ($ * $) operation 
is defined on the $ D $-dimensional spacetime manifold (see, e.g. [4-10]). 
It is straightforward to check that $\delta A^{(1)} = \pm * d * (dx^\mu\, A_\mu ) 
= \pm\, (\partial\cdot\, A ) $ is a zero-form.
We have a freedom to add/subtract a zero-form scalar field ${\phi}$ to this gauge-fixing term
(with the proper mass dimension). This is why, in this total gauge-fixing term $[- (\,
\partial\cdot A + m\, \phi)^2 / \,2]$, we have the presence of $m\,\phi $, too. We shall discuss this issue of adding the extra term (i.e. $m\,\phi$) later in detail (see, e.g. Sec. 3). 
Under the on-shell nilpotent (anti-)BRST symmetry transformations $s_{(a)b}$, 
the Lagrangian density (5) transforms to the total spacetime derivatives:
\begin{eqnarray}
s_{ab}\, {\cal L}_B &=& -\,{\partial_\mu\, \Bigl[\, (\partial\cdot A 
+ m \, \phi)\,\partial ^\mu\, \bar C \,\Bigr]}, \nonumber\\ 
s_{b}\, {\cal L}_B &=& -\,{\partial_\mu\, \Bigl[ \,(\partial\cdot A 
+ m \, \phi)\,\partial ^\mu\, C \,\Bigr]}.
\end{eqnarray}
Thus, the action integral $ S= \int d^{D-1}x\, ({\cal L}_B) $ remains invariant under the (anti-)BRST symmetry transformations  $ s_{(a)b}$. The above continuous symmetry transformations, according to Noether's theorem, lead to the following conserved charges $ Q_{(a)b}$, namely;
\begin{eqnarray}
Q_{ab} = \int d^{D-1}x \,\Bigl[ \,
 \partial_0 \{(\partial\cdot A) + m \,\phi \} \,\bar C -\,\{ (\partial\cdot A + m\,\phi )\} \dot{\bar C} \,\Bigr ],\nonumber\\
Q_{b} = \int d^{D-1}x\, \Bigl[ \, \partial_0\{ (\partial\cdot A) + m\, \phi \} \,C -\{ (\partial\cdot A + m\,\phi ) \}\, \dot C \,\Bigr ],
\end{eqnarray}
which turn out to be the generators for the above on-shell nilpotent symmetry transformations $ s_{(a)b}$ because $s_{(a)b}\, \Psi = \pm\, i \,[\Psi,\, Q_{(a)b}]_{\pm} $ for the generic field $ \Psi\,\equiv A_\mu,\, \phi,\, C, \,\bar C $ of the theory, described by the Lagrangian density (5). Here the ($\pm$) signs, as the subscript on the square bracket, correspond to the (anti)commutator for $\Psi$ being (fermionic)bosonic.

We wrap up this section with the following remarks. First, our whole argument is valid in 
any arbitrary dimension of spacetime. Second, the (anti-)BRST symmetry transformations are
 nilpotent of order two $(s_{(a)b}^2 = 0)$ when we use the equations of motion: 
$(\Box + m^2)\,C = 0, (\Box + m^2)\,\bar C = 0$. Third, these transformations are absolutely 
anticommuting $(s_b\, s_{ab} + s_{ab}\,s_b = 0) $ in their operator form when we exploit the 
on-shell conditions $(\Box + m^2)\,C = 0, (\Box + m^2)\,\bar C =0 $. Finally, the above nilpotent 
symmetries have their mathematical origin in the exterior derivative $d$ (with $d\, ^2 = 0$)
 because the curvature term $ F_{\mu\nu}= \partial_\mu\, A_\nu - \partial_\nu\, A_\mu $
[which remains invariant under $s_{(a)b}$ (cf. 4 )] is generated from 
$d\,A^{(1)} = [(dx^\mu \wedge dx^\nu)/2!]\, F_{\mu\nu}$. Physically, it is obvious
 to note that the kinetic term (for the gauge field) remains invariant under $s_{(a)b}$
 (which is a characteristic feature of (anti-)BRST symmetries). We further re-emphasize that, even though,  the
 mathematical origin for the existence of (anti-)BRST symmetries is encoded in the
 exterior derivative $d = dx^\mu \partial_\mu$, {\it only} one of these symmetries could
 be identified with $d$ because $s_b$ and $s_{ab}$ respect the absolute anticommutatvity 
 property and, hence, they are linearly independent of each-other.

\section{On-shell nilpotent (anti-)co-BRST symmetries}

Let us begin with the two (1+1)-dimensional (2D) version of 
the (anti-)BRST invariant Lagrangian density (5). This can be expressed as\footnote{ We adopt the conventions and notations such that the 2D background flat Minkowski spacetime has the metric $\eta_{\mu\nu} = $ diag $(+1,-1)$ so that 
$ A\cdot B = \eta^{\mu\nu} A_\mu B_\nu \equiv A_\mu B^\mu = A_0 B_0  -A_1 B_1$ is the dot product between two non-null vectors $A_\mu$ and $B_\mu$. We have also the Levi-Civita tensor $\varepsilon_{\mu\nu}$ with the convention $\varepsilon_{01} = + 1 = \varepsilon^{10},\, 
\varepsilon^{\mu\lambda}\, \varepsilon_{\lambda\nu} =  \delta^\mu_\nu$, etc.,  and the d'Alembertian operator in our theory is:
$\Box = {\partial_0}^2 - {\partial_1}^2 $.}:
\begin{eqnarray}
{\cal L}_b &=& \frac {1}{2}\, E^2 + \frac {m^2}{2} A_\mu\, A^ \mu + \frac {1}{2}\, \partial_\mu\, \phi\, \partial^\mu\, \phi - m A_\mu \,\partial^\mu\, \phi \nonumber\\
 &-&\frac {1}{2}\,(\partial\cdot A + m\,\phi)^2 -
 i\,\partial_\mu\,\bar C\,\partial^\mu\, C + \, i\,m^2 \,\bar C\, C,
 \end{eqnarray}
where $F_{\mu\nu} $ tensor has only electric field $(E = F_{01} \equiv -\, \varepsilon^{\mu\nu} \,\partial_\mu\, A_\nu) $ as the non-vanishing component and there is {\it no} magnetic field ($B$) in the theory.
It is obvious that, the existing $E$ field (with {\it one} component) is a {\it pseudo-scalar} because it changes sign under parity.

The kinetic term $(E^2/2)$ of the (anti-)BRST invariant Lagrangian density (8)  can be generalized. 
To achieve this goal in a symmetric fashion, one has to note that,  we have the following expression for the Hodge dual
of $F^{(2)}$, in 2D spacetime, namely; 
\begin{eqnarray}
*\; F^{(2)} = * \; \Bigl (\frac {d x^\mu \wedge d x^\nu}{2!} \Bigr )\; F_{\mu\nu} \equiv \varepsilon^{\mu\nu}
\; \partial_\mu \, A_\nu = -\, E,
\end{eqnarray}
where $*$ is the Hodge duality operation. We observe that the electric field $E$
(originating from $F^{(2)} = [(dx^\mu \wedge dx^\nu)/2!]\, F_{\mu\nu}$ in the 2D spacetime)  is 
an anti-self-dual field  and it is a 0-form pseudo-scalar. Thus, there is a room for adding/subtracting a 0-form 
pseudo-scalar field $(\tilde\phi)$, with a proper mass dimension, in the expression for the kinetic term $E^2/2$
of the Lagrangian density (8) in two (1+1)-dimensions of spacetime.

Following the above arguments, it can be seen that the above Lagrangian density (8) can be generalized 
symmetrically into the following couple of  forms:
\begin{eqnarray} 
{\cal L}_{(b_1)} &=& \frac {1}{2}\, {(E - m\,\tilde\phi)}^2 + m\, E\,\tilde\phi -\frac {1}{2}\,\partial_\mu \,\tilde\phi\,\,\partial^\mu\,\tilde\phi +\frac {m^2}{2} A_\mu\, A^ \mu + \frac {1}{2}\, \partial_\mu\, \phi\, \partial^\mu\, \phi 
\nonumber\\ &-& m A_\mu \,\partial^\mu\, \phi
 -\frac {1}{2}\,(\partial\cdot A + m\,\phi)^2 - 
 i\,\partial_\mu\,\bar C\,\partial^\mu\, C + \, i\,m^2 \,\bar C\, C,
\end{eqnarray}
\begin{eqnarray} 
{\cal L}_{(b_2)} &=& \frac {1}{2}\, {(E + m\,\tilde\phi)}^2 - m\, E\,\tilde\phi - \frac {1}{2}\,\partial_\mu \,\tilde\phi\,\,\partial^\mu\,\tilde\phi 
+ \frac {m^2}{2} A_\mu\, A^ \mu + \frac {1}{2}\, \partial_\mu\, \phi\, \partial^\mu\, \phi 
\nonumber\\ &+& m A_\mu \,\partial^\mu\, \phi
 -\frac {1}{2}\,(\partial\cdot A - m\,\phi)^2 -
 i\,\partial_\mu\,\bar C\,\partial^\mu\, C + \, i\,m^2 \,\bar C\, C.
\end{eqnarray}
The above Lagrangian densities\footnote{In fact, it is straightforward to check that both the Lagrangian densities differ by a total spacetime derivative term. Thus, they are {\it equivalent} as far as the dynamics 
and symmetries are concerned.} show, in explicit form, the addition/subtraction of the (pseudo-)scalar fields in the kinetic and gauge-fixing terms of the present 
massive gauge theory\footnote{A close look at the transformations $A_\mu \to A_\mu 
\mp (1/m)\, \partial_\mu \phi$ [used in the Stueckelberg formalism (keeping 
$F_{\mu\nu} \to F_{\mu\nu}$)] and duality transformations (cf. (29) below) ensure the
appearance of terms like $(E \,\mp\, m\, \tilde \phi)$ and 
$(\partial \cdot A\, \pm \,m\, \phi)$ 
in the Lagrangian densities (10) and (11) for our present 2D theory. Mathematical origin for these terms, based on 
the methodology of differential geometry, has already been explained in 
the main body of our text. Thus, Lagrangian densities (10) and (11) are very 
appropriate.}.
We shall concentrate here only on ${\cal L}_{(b_1)}$. However, in our Appendix C, we shall 
mention a few things about symmetries of the Lagrangian density ${\cal L}_{(b_2)}$, too.
The above Lagrangian density ${\cal L}_{(b_1)}$ respects the following on-shell
[$(\Box + m^2 )\,C = 0,\,(\Box + m^2 )\,\bar C = 0$] nilpotent  ($s_{(a)d} ^2 = 0 $) (anti-)co-BRST symmetry transformations:
\begin{eqnarray}
&& s_{ad}\, A_\mu = -\, \varepsilon _{\mu\nu}\,\partial^\nu \,C,\qquad
 \,s_{ad} \,C = 0,\qquad\qquad \,\,\,s_{ad}\, \bar C = + \,i\,(E - m\,\tilde\phi ),\nonumber\\ 
&& s_{ad}\, E = \Box \,C,\qquad s_{ad}\,(\partial \cdot A + m \,\phi) = 0,
 \quad s_{ad}\, \phi = 0,\quad s_{ad}\, \tilde\phi= - m\, C,\nonumber\\ 
&& s_d\, A_\mu = -\, \varepsilon _{\mu\nu}\,\partial^\nu\, \bar C,\qquad\quad
s_d \,\bar C = 0,\qquad\qquad\,\,\,s_d \,C = -\, i\,(E - m\,\tilde\phi ),\nonumber\\  
&& s_d\,E = \Box\, \bar C, \quad\,s_d \,(\partial \cdot A + m \,\phi) = 0,
 \qquad s_d\, \phi = 0, \qquad s_d\, \tilde\phi= - m\,\bar C, 
\end{eqnarray}
We note that the gauge-fixing term, owing its origin to the 
nilpotent ($\delta^2 = 0$) co-exterior derivative ($ \delta = \pm * d * $ ), remains invariant under the (anti-)co-BRST symmetry transformations $s_{(a)d}$. This is why we have christened, the symmetry transformations in (12), as the (anti-)co-BRST [or (anti-)dual-BRST] symmetries. These should be contrasted with the (anti-)BRST symmetry transformations $s_{(a)b}$, under which, the total kinetic term (owing its origin to the exterior derivative $d = dx^\mu \partial_\mu$) remains invariant. For the above modified Lagrangian density (10), the latter transformations (i.e. $s_{(a)b}$) are 
\begin{eqnarray}
s_{ab}\, \tilde\phi = 0,\qquad \quad s_{ab}\,E = 0, \qquad \quad
s_b\tilde\phi=0,\qquad \quad s_b E = 0, 
\end{eqnarray}
in addition to the ones listed in (4). Under the total (anti-)BRST symmetry transformations (4) and (13), the Lagrangian density (10) transforms to the total spacetime derivatives exactly in the same way as given in (6) as there is no contribution from the new terms.

We observe that, under the (anti-)co-BRST symmetry transformations (12), the Lagrangian density (10) transforms to the total spacetime derivative as given below: 
\begin{eqnarray}
s_{ad}\, {\cal L}_{{(b_1)}} &=& \partial_\mu \,\Bigl [ E\,\partial^\mu \,C + m \,\varepsilon^{\mu\nu} \,(m \,A_\nu \,C + \phi\, \partial_\nu\, C ) \Bigr ], \nonumber\\
s_d\,{\cal L}_{{(b_1)}} &=& \partial_\mu \,\Bigl [ E\,\partial^\mu \,\bar C + m \,\varepsilon^{\mu\nu}\, (m \,A _\nu \,\bar C + \phi\, \partial_\nu\,\bar C ) \Bigr ], 
\end{eqnarray}
which demonstrates that the action integral $S = \int dx \, {\cal L}_{(b_1)}$ remains invariant under the transformations (12). Applying Noether's theorem, we obtain the following conserved charges corresponding to the continuous symmetry transformations (12), namely;
\begin{eqnarray}
Q_{d} &=& \int \, dx\, \Bigl [\,m\,(\dot{\tilde\phi}\,C - \tilde\phi\, \dot{\bar C}) + (E\,\dot {\bar C} 
- \dot E\, {\bar C}) \Bigr ], \nonumber\\
&\equiv& \int \,dx\, 
\Bigl [\; (\,E - m\, \tilde\phi )\,\dot {\bar C} - (\dot E - m\,\dot{\tilde\phi})\,\bar C \Bigr ], \nonumber\\ 
Q_{ad} &=& \int\, dx \;\Bigl [\, ( E\,- m\,\tilde\phi )\,\dot C - (\dot E\, - m\,
\dot {\tilde\phi)}\,C],
\end{eqnarray}
which are generators for the transformations (12). The above charges $ Q_{r} = \int dx\, J^0_{r} \,(r = d, ad) $ have been derived from the following Noether currents: 
\begin{eqnarray}
 J^\mu_{(d)} &=&  E\,\partial^\mu \,\bar C + \varepsilon^{\mu\nu}\,{(\partial\cdot A)}\,\partial_\nu\, \bar C - m^2\,\varepsilon^{\mu\nu}\,A_\nu\,\bar C \nonumber\\ &+& m\,( \bar C\,\partial^\mu \,\tilde\phi - \tilde\phi\,\partial^\mu \bar C),\nonumber\\
 J^\mu_{(ad)} &=&  E\,\partial^\mu\, C + \varepsilon ^{\mu\nu}\,{( \partial\cdot A)}\, \partial_\nu \,C - m^2\,\varepsilon^{\mu\nu} \,A_\nu \,C,\nonumber\\
 &+& m\,( C\,\partial^\mu \,\tilde\phi - \tilde\phi\,\partial^\mu C),
\end{eqnarray}
In the proof of the conservation law $\partial_\mu J^\mu_{(a)d}=0$, we have to exploit the following Euler-Lagrange equations of motion for the Lagrangian density (10), namely;
\begin{eqnarray}
&&\varepsilon^{\mu\nu}\,\partial_\nu E = \partial^\mu\,(\partial\cdot A) + m^2 A^\mu, \qquad\qquad\qquad (\Box + m^2) A_\mu=0,\nonumber\\
&&(\Box + m^2)\,\phi = 0,\qquad\; (\Box+m^2)\,\tilde\phi=0,\qquad \;(\Box +m^2) (\partial\cdot A)=0,\nonumber\\
&&(\Box +m^2) \;E = 0,\;\qquad (\Box+ m^2) \,C = 0,\;\qquad (\Box + m^2)\, \bar C = 0.
\end{eqnarray}
It should be noted that $ ( \Box+ m^2 )\, A_\mu\,=\,0$, $( \Box+ m^2 )\, E\,=\,0$ ,  $( \Box+ m^2 )\,(\partial\cdot A)\,=\,0$ have emerged out from the single equation $ \varepsilon^{\mu\nu}\,\partial_\nu\, E= \partial^\mu \,(\partial\cdot A) + m^2 A^\mu$.

We wrap up this section with the following comments. First, to obtain the on-shell $\Bigl [( \Box+ m^2 ) C=0,( \Box+ m^2 ) \bar C=0 \Bigr]$ nilpotent $(s^2_{(a)d}=0)$  (anti-)co-BRST symmetries, we have invoked a pseudo-scalar field $(\tilde\phi)$ in the theory.  Second, it can be explicitly checked that $(s_d\,s_{ad}\,+\,s_{ad}\,s_d\,=\,0$)
when we use the equations of motion: $ ( \Box+ m^2 )\, C=0\,$ and $ ( \Box+ m^2 )\, \bar C=0$. Third, the on-shell [$ ( \Box+ m^2 )\, C=0,\,( \Box+ m^2 )\, \bar C=0 $] nilpotency $(Q^2_{(a)d}=0)$ and anticommutativity ($Q_d\,Q_{ad} + Q_{ad}\,Q_d = 0$) of the (anti-)co-BRST charges $Q_{(a)d}$ can 
{\it also} be checked by using the following formula for the generators, namely:
\begin{eqnarray}
&& s_d Q_d = i \,\{ Q_d,\, Q_d\}\,=\,0,\qquad\qquad s_{ad} Q_{ad}= i \,\{Q_{ad},Q_{ad}\}\,=\,0,\nonumber\\
&& s_d Q_{ad} = i \,\{Q_{ad},\, Q_d\}\,=\,0,\qquad\quad s_{ad} Q_d = i \,\{Q_d,\, Q_{ad}\}\,=\,0,
\end{eqnarray}
where the l.h.s. of the above relations can be calculated easily from (12) and (15). Fourth, the Lagrangian
density (8) does {\it not} respect any duality symmetry as the equivalent Lagrangian densities (10) and (11)
do [cf. (29)]. However, the former Lagrangian density respects a couple of supersymmetric type continuous
symmetry transformations and a bosonic symmetry. These symmetries have been briefly discussed in our
Appendices A and B.  
Finally, the (pseudo-)scalar fields $(\tilde \phi)\phi$ have been 
added/subtracted in a very symmetrical fashion to the kinetic and gauge-fixing terms of the theory. These are ``dual'' to each-other as would become 
clear in equation (29) (see below).

\section{Bosonic symmetry and its uniqueness}

In our present 2D theory, so far, we have discussed four nilpotent $(s_{(a)b}^2 = 0,\, s_{(a)d}^2 = 0)$ symmetries which are $s_{(a)b}$ and $s_{(a)d}$. 
It turns out that we have the validity of the following:
\begin{eqnarray}
\{s_b, s_{ab}\} =0,\qquad \{s_b, s_{ad}\} = 0,\qquad \{s_d, s_{ab}\} = 0, \qquad\{s_d, s_{ad}\} = 0,
\end{eqnarray}
where the on-shell conditions, from the equation 
of motion (17), have to be exploited for their proof. The following 
{\it unique} bosonic symmetry ($s_w$), emerging due to the anticommutatation of the 
{\it two} fermionic symmetries, is defined as:
\begin{eqnarray}
\,s_w\,=\,\{s_b,\, s_d \}\,=\,-\,\{s_{ad},s_{ab}\},
\end{eqnarray}
where the nilpotent transformations $s_{(a)b}$ and $s_{(a)d}$ are listed in (4) and (12).

The infinitesimal and continuous symmetry transformations ($s_w$) for the individual relevant fields 
(modulo a factor of i) are as follows:
\begin{eqnarray}
&& s_w\,A_\mu\;=\;\varepsilon_{\mu\nu}\, (\Box\,A^\nu\,+\,m \,\partial^\nu\,\phi ) + \, m\, \partial_\mu\,\tilde\phi,\qquad s_w\,(C,\bar C)\,=\,0, \nonumber\\
&& s_w\,\phi\,= \,-\,m\,(E - m\,\tilde \phi), \quad \qquad \qquad \qquad \qquad  s_w\,\tilde\phi = \, m\, 
(\partial \cdot A + \, m\,\phi),\nonumber\\
&& s_w\,E\,=\,-\,\Box\,(\partial\cdot A + \,m\,\phi),\qquad \qquad\quad
\qquad s_w\,(\partial\cdot A)\,=\,-\,\Box\, (E - m\tilde\phi).
\end{eqnarray}
The noteworthy point, at this stage, is the fact that the (anti-)ghost fields $(\bar C) C$ do not transform under $s_w$. Thus, one of the decisive features of $s_w$ is the observation that the ghost part of the Lagrangian density does not 
change {\it at all} under the local and continuous bosonic symmetry transformations ($s_w$).

 Under the above infinitesimal continuous transformations, the Lagrangian density ${\cal L}_{(b_1)}$ transforms 
to a total spacetime derivative, as:
\begin{eqnarray}
s_w {\cal L}_{(b_1)} &=& \partial_\mu  \Bigl[(\partial \cdot A + m\, \phi)\partial^\mu \,(E - m\,\tilde\phi ) - \,( E - m\,\tilde\phi)\,\partial^\mu\,(\partial\cdot\ A + m\,\phi)\nonumber\\ &+&  \,m^2\, (A^\mu E + \varepsilon^{\rho\sigma} A_\rho\,\partial^\mu A_\sigma) - \, m\, \varepsilon^{\mu\nu}\,(\phi\,\Box\, A_\nu + m^2\, A_\nu\, \phi) \nonumber\\
&-& \, m\,\{\phi\,\partial^\mu\, E +\tilde\phi\, \partial^\mu \,(\partial\cdot A) \}
- \, m^2\, \tilde\phi\, \partial^\mu\,\phi \Bigr].
\end{eqnarray}
As a consequence, the action integral $S = \int dx \, ({\cal L}_{(b_1)})$ remains invariant. According to Noether's theorem, the above continuous symmetry transformations lead to the derivation of conserved charge,
for our present 2D theory, as:
\begin{eqnarray}
Q_w = \int dx\,\Bigl[(\partial\cdot A + m\, \phi)\,(\dot E - m\, \dot {\tilde\phi}) 
- \partial_0\, \{(\partial\cdot A + m\, \phi) \} \,(E - m\, \tilde\phi )\Bigr].
\end{eqnarray}
The above charge is the generator of the continuous and infinitesimal transformations $(s_w)$.

We close this section with the remark that, by exploiting the definition of a generator, the following anticommutators (and transformations
on the appropriate charges), namely;
\begin{eqnarray}
&& s_b\, Q_d = i\, \{ Q_d,\, Q_b\} = i\, Q_w, \qquad\qquad\, s_d\, Q_b = i\, \{ Q_b,\, Q_d\} = i\, Q_w, \nonumber\\
&& s_{ad}\, Q_{ab} = i\, \{ Q_{ab},\, Q_{ad}\} = -\, i\, Q_w, \qquad s_{ab}\, Q_{ad} = i\, \{ Q_{ad},\, Q_{ab} \} = -\, i\, Q_w,
\end{eqnarray}
{\it also} produce the expression for $Q_w$. We lay emphasis on the fact that, in the calculation of (24), we have to use the transformations (4) and (12) as well as the expressions for the charges $Q_{(a)b}$  and $Q_{(a)d}$ that are quoted in the equations (7) and (15). This method of calculation 
of the bosonic charge ($Q_w = \{ Q_b, Q_d \} \equiv
- \, \{Q_{ad}, Q_{ab} \} $) is simpler than the usual application of the Noether's theorem
 where the algebra is quite involved. It can be clearly checked, however, that the expression
 for $Q_w$ (cf. (23)) is found to be exactly the same when we use the Noether's theorem for the
 computation of $Q_w$.

\section{Ghost-scale and discrete symmetries}

It is straightforward to note that our present theory is endowed with the ghost-scale symmetry transformations 
(on the basic fields and their composites) as illustrated below:
\begin{eqnarray}
\Psi\rightarrow e^{0\,.\,\Lambda}\, \Psi, \qquad C\rightarrow e^{+\,\Lambda}\,C,\qquad \bar C\rightarrow e^{-\,\Lambda}\, \bar C, 
 \end{eqnarray}
where $\Lambda$ is a global (spacetime independent) parameter and its coefficients denote the ghost number for a given field. The generic field $\Psi = \phi, A_\mu, \tilde\phi, E, (\partial\cdot A)$ is endowed with the ghost number 
equal to zero and (anti-)ghost fields $(\bar C)C$ carry the ghost numbers $(\mp)\, 1$, respectively. The infinitesimal version of the ghost-scale symmetry
transformations (25), denoted by $(s_g)$, is as follows for the relevant fields of the theory, namely;
\begin{eqnarray}
s_g\Psi = 0,\quad\qquad s_g C = + \,C,\quad\qquad s_g \bar C = - \,\bar C,
\end{eqnarray}
where the generic field $\Psi = A_\mu, \phi, \tilde\phi, (\partial\cdot A), E$ and we have chosen $\Lambda = 1$ for the sake of brevity. We can readily 
verify, using the idea of a generator, that:
\begin{eqnarray}
&& s_g Q_b = -\, i\, \Bigl[Q_b, Q_g \Bigr] = +\,Q_b\,\,\quad\Longrightarrow i\, \Bigl[Q_g, Q_b\Bigr] = +\, Q_b,\nonumber\\
&& s_g Q_{ab} = -\, i\, \Bigl[Q_{ab}, Q_g \Bigr] = -\, Q_{ab}\,\Longrightarrow i \Bigl[Q_g, Q_{ab}\Bigr] = -\, Q_{ab},\nonumber\\ && s_g Q_d = -\, i\, \Bigl[Q_d, Q_g \Bigr] = -\, Q_d\,\quad\Longrightarrow i \Bigl[Q_g, Q_d\Bigr] = -\, Q_d,\nonumber\\
&& s_g Q_{ad} = -\, i\, \Bigl[Q_{ad}, Q_g \Bigr] = +\, Q_{ad} \Longrightarrow i \Bigl[Q_g, Q_{ad}\Bigr] = +\, Q_{ad},\nonumber\\
&& s_g Q_w = -\, i\, \Bigl[Q_w, Q_g \Bigr] = \,0 \,\,\qquad\Longrightarrow \,i \,\Bigl[Q_g, Q_w\Bigr] = \,0,
\end{eqnarray}
where the ghost charge $Q_g$ is the generator for the infinitesimal transformations $(s_g)$, quoted in 
the equation (26). Its explicit expression,
in terms of the (anti-)ghost fields,  is 
\begin{eqnarray}
Q_g = i \int d x \,\Bigl[\bar C \; \dot C - \dot{\bar C} \; C \Bigr], 
\end{eqnarray}
which is derived from the conserved $(\partial_\mu J^\mu _{(g)} \,=\, 0)$ Noether current $J^\mu _{(g)} = i \Bigl[ \bar C \,\partial^\mu C - (\partial^\mu \,\bar C)C)\Bigr]$.
The l.h.s. of every one of 
the above set of equations (27) is very straightforward to calculate by taking the help of (26), (7), (15) and (23).

There exists a set of beautiful discrete symmetries in the theory. For instance, it can be readily checked that, under the following discrete symmetry transformations, namely;
\begin{eqnarray}
&& A_\mu \rightarrow \pm\, i\, \varepsilon_{\mu\nu}\, A^\nu, \quad\qquad \qquad \phi \rightarrow \pm\, i \,\tilde\phi,
\qquad\qquad \qquad \tilde\phi\rightarrow \pm\, i \,\phi, \nonumber\\
&& C\rightarrow \mp\, i\, \,\bar C,\qquad \;\bar C \rightarrow \mp\, i\, C,\qquad\; (\partial\cdot A) \rightarrow \mp \,i \,E,\qquad\; E\rightarrow \mp\, i \,(\partial\cdot A),
\end{eqnarray}
the Lagrangian densities (10) and (11) remain invariant. We shall see, later on, that the above discrete symmetry transformations play very important roles in establishing the connection between the continuous symmetry transformations and the de Rham cohomological operations of differential geometry. In principle, we can have many discrete symmetry transformations in the theory. However, we have quoted only $two$, in the above, which are very useful to us in our subsequent discussions.

We dwell a bit on the origin of transformations (29) which are symmetry transformations for the Lagrangian  densities (10) and (11). We note that the 2D 
self-duality condition, for the 1-form gauge connection, is as follows:
\begin{eqnarray}
*\, A^{(1)} = *\, (dx^\mu A_\mu) = dx^\mu \tilde A _\mu ,
\end{eqnarray}
where  $\tilde A_\mu = - \varepsilon_{\mu\nu}\, A^\nu$.
This relation leads us to a clue to find out the discrete symmetry transformations (29). 
In fact, it is the transformations $A_\mu \rightarrow \pm\, i\,\varepsilon_{\mu\nu}\, A^{\nu}$, 
owing their origin to (30), that are at the heart of the 
transformations (29). To be more precise, it can be seen that, for the 
following gauge-fixed Lagrangian density (in the Feynman gauge) of a free 2D Proca theory 
(without the inclusion of Stueckelberg field $\phi$): 
\begin{eqnarray} 
{\cal L}_{(free)} = \frac {1} {2}\, E^2 - \frac{1}{2}\, (\partial\cdot A)^2 + \frac{m^2}{2}\, A_\mu\, A^\mu ,
\end{eqnarray}
the discrete symmetry transformations are: $A_\mu \rightarrow \pm \,i\,\varepsilon_{\mu\nu}\, A^{\nu}$ (see e.g. [20,21] for more details). The transformations for 
the other fields (e.g. $\phi, \tilde\phi, C, \bar C)$ are motivated from the starting transformations $A_\mu \rightarrow \pm\, i\,\varepsilon_{\mu\nu}\, A^{\nu}$ so as to have a {\it perfect} symmetry for the Lagrangian densities (10) and (11). 
It is worthwhile to point out that $C\rightarrow \mp\, i \,\bar C,\, \bar C \rightarrow \mp\, i\, C$  transformations are {\it also} the discrete symmetry transformations for the ghost 
part of the Lagrangian densities (10) and (11). Thus, the discrete symmetry transformations (29) 
are very appropriate
because they incorporate all the essential ingredients of our theory.

\section{Extended BRST algebra: cohomological aspects}

We have discussed, so far, a set of {\it six } local and continuous symmetry transformations (i.e. $s_{(a)b},\,  s_{(a)d},\,    s_g,\,  s_w  $) and a set of $two$ discrete symmetry transformations (29). In their operator form, the continuous symmetry transformations obey the following algebra:
\begin{eqnarray}
&& s_{(a)b}^2 = 0,\;\qquad s_{(a)d}^2 = 0,\;\; \qquad \{s_b, s_{ab} \} = 0, \;\;\qquad \{s_d, s_{ad} \} = 0,\nonumber\\ 
&&\{s_b, s_{ad} \} = 0,\quad \{s_d, s_{ab} \} = 0,\;\;\quad \{s_b, s_{d} \} = s_w = \, - \, \{s_{ab}, s_{ad} \}, \nonumber\\
&& [s_w,\, s_r] = 0, \qquad r = b,\, ab, \,d,\, ad,\, g,\, \qquad [s_g,\, s_d] = - s_d,\nonumber\\
&& [s_g,\, s_b] = s_b,\qquad \, [s_g,\, s_{ab}] = -\, s_{ab}, \,\,\,\,\qquad \, [s_g,\, s_{ad}] = s_{ad} .
\end{eqnarray}
Thus, we note that the transformation
$s_w$ is the Casimir operator for the whole algebra. Exploiting the definition of a generator for a given transformation, we can replicate the above 
algebra (32), in the language of conserved charges of the theory, as follows:
\begin{eqnarray} 
&& Q_{(a)b}^2 = 0,\,\qquad\{Q_b, Q_{ab} \} = 0,\,\qquad  \{Q_d, Q_{ad} \} = 0,\qquad \{Q_b, Q_{ad} \} = 0,\nonumber\\
&& \{Q_d, Q_{ab} \} = 0,\qquad \qquad \;\{Q_b, Q_{d} \} \;= \;Q_w = \, - \, \{Q_{ab}, Q_{ad}\} ,\nonumber\\
&& [Q_w,\, Q_r] = 0,\qquad \quad r = b,\, ab, \,d,\, ad, \,g, \qquad \quad i\,[Q_g,\, Q_d] = - \,Q_d,\nonumber\\
&& i\,[Q_g,\, Q_b] = Q_b,\qquad\quad \, i\,[Q_g,\, Q_{ab}] = -\, Q_{ab},\qquad\quad i\,[Q_g,\, Q_{ad}] = Q_{ad}, 
\end{eqnarray}
which shows that $Q_w$ is the Casimir operator for the whole algebra. It is worthwhile to point out that the 
algebras (32) and (33) are satisfied 
{\it only} on the on-shell (where the Euler-Lagrange equations of motion (17) are satisfied).

A close look at the equations (32) and (33) demonstrates that these extended algebras  are very similar to the following algebra satisfied by the de Rham cohomological operators ($ d,\, \delta,\, \Delta$) of differential geometry [4-10], namely; 
\begin{eqnarray}
&& d^2= 0,\qquad \delta^2= 0,\qquad \{d,\,\delta\}=\,\Delta,\qquad \Bigl[\Delta,\,\delta\Bigr]= 0,\nonumber\\
&& \Delta\, = (d\,+\,\delta)^2\,=\, d\,\delta +\, \delta\, d,\qquad \quad \Bigl[\Delta,\,d\,\bigr]=\, 0,
\end{eqnarray}
which shows that the Laplacian operator $\Delta\,= \,\delta\, d+\,d\,\delta \,$  is the Casimir operator for the whole algebra (34). As far as the algebraic structures (32), (33) and (34) are concerned, it is evident that there is an analogy between the transformation operators (and corresponding conserved charges) and the cohomological operators of differential geometry. These are: $(s_b,\,s_{ab})\rightarrow d, \,(s_d,\,s_{ad})\rightarrow \,\delta,\;s_w\,=\, \{s_d,\,s_b\}\,=\,\Delta\,\equiv \,-\{s_{ab},\,s_{ad}\} $ at the level of transformation operators and at the level of conserved charges, we have:
$(Q_b,\,Q_{ab})\,\rightarrow d,\,(Q_d,\,Q_{ad})\rightarrow \delta $ and $Q_w\, = \,\{Q_b,\,Q_d \}\, \equiv -\,\{Q_{ab},\,Q_{ad}\}\,\rightarrow \,\Delta$.

The identifications made, in the above, are $not$ complete as yet. There are missing points that we have to 
answer and fulfill before the  sanctity of the
above identifications could be justified and put on the firmer footings. The 
{\it first} issue is, as we know, the 
nilpotent exterior derivative $(d)$ and co-exterior derivative ($\delta$) are connected by the relations:
\begin{eqnarray}
\delta=\,-\,*\,d\,\,*,\qquad\qquad d^2=0, \qquad\qquad{\delta}^2=0,\
\end{eqnarray}
where ($*$) is the Hodge duality operation and the minus sign, in the above relationship, is due to the $even$ dimensionality of the spacetime manifold. It is very interesting to state that the relation (35) is satisfied, in our theory, by the following operator relationships:
\begin{eqnarray}
s_{(a)d}=\,- \,*\,s_{(a)b}\, \,*\,,\qquad\qquad s_{(a)b}^2=\,0, \qquad\qquad s_{(a)d}^2=\,0,
\end{eqnarray}
where $s_{(a)b}$ and $s_{(a)d}$ are the continuous symmetry transformations (4) (as well as (13))
and (12), respectively, and $(*)$ corresponds to a couple of discrete symmetry transformations quoted in (29). It is because of the dimensionality of our theory that there exists an inverse relationship between $s_{(a)b}$ and $s_{(a)d}$ as given by the following relationship:
\begin{eqnarray}
s_{(a)b}\, =\, -\, *\, s_{(a)d}\,* ,\qquad \qquad s_{(a)b}^2\, =\,0, \qquad \qquad s_{(a)d}^2\, =\, 0, 
\end{eqnarray}
where the interplay between the continuous symmetry
transformations ($s_{(a)b}, s_{(a)d}$) and the discrete symmetries (29) play a clinching and decisive role. The minus sign, in (36) and (37), is decided by the following relationship for the generic field:
\begin{eqnarray}
*\, ( *\, \Psi) = \,-\, \Psi,\qquad\qquad \Psi = A_\mu,\, \phi,\, \tilde\phi,\, C,\, \bar C,\, (\partial\cdot A),\, E,
\end{eqnarray}
which is true for any duality invariant theory [29]. 
In the above, the l.h.s. denotes the two successive operations of the discrete symmetry transformations (29).

In differential geometry, we know that the operation of an exterior derivative on a form ($f_n$) of degree $n$, raises the degree of the form by one (i.e. $ d\, f_n \sim f_{n+1}$). On the contrary, when the co-exterior derivative $\delta$ acts on a form ($f_n$) of degree $n$, it lowers the degree of the form by one (i.e. $\delta f_n \sim f_{n-1}$ ). Furthermore, as we know, the degree of a form remains intact when it is acted upon by the Laplacian operator $\Delta$ (i.e. $\Delta\, f_n \sim f_n$ ). Thus, the {\it second} issue to be resolved is that
we have to capture these properties in the language of symmetry operators and corresponding conserved charges. In fact, it is the equation (33) that plays a clinching and decisive role here in capturing the above property in a cogent and convincing fashion. We discuss this analogy, using the algebra (33), from now on.

We observe, from the above arguments, that the operation of $d\,(\delta)$ on the form $(f_n)$ of degree $n$, is like the properties associated with the $ladder$ operators. To capture the latter property, we define a state $\mid \chi\rangle_n$, with the ghost number $n$, in the quantum Hilbert space of states of a BRST invariant theory as follows:
\begin{eqnarray}
i\, Q_g \,| \chi\rangle_n\, =\, n\, |\chi\rangle_n,
\end{eqnarray}
where $Q_g$ is the ghost charge (28). Exploiting the strength of the algebra (33), we observe the following interesting relations amongst the charges, namely;
\begin{eqnarray}
&&i\, Q_g\, Q_b\,  |\chi\rangle_n\,=\, (n+1)\, Q_b\, | \chi\rangle_n,  \quad i\, Q_g\,Q_{ad}\,| \chi\rangle_n\,=\, (n+1)\,Q_{ad}|\chi\rangle_n, \nonumber\\
&&i\, Q_g\, Q_d\,  |\chi\rangle_n\,=\, (n-1)\, Q_d\, | \chi\rangle_n,  \quad i\, Q_g\, Q_{ab}\,|\chi\rangle_n\,=\, (n-1)\,Q_{ab}|\chi\rangle_n, \nonumber\\
&&i  i\,Q_g\,Q_w\,|\chi\rangle_n\,=\,n\,Q_w\,|\chi\rangle_n.
\end{eqnarray}
A close look at the above relations justify that the states $Q_b\,|\chi\rangle_n$, $Q_d\,|\chi\rangle_n$ and $Q_w\,|\chi\rangle_n$ have the ghost numbers $(n+1),\,(n-1)$ and $n$, respectively. The same ghost numbers are associated with the states $Q_{ad}\, | \chi\rangle_n,\, Q_{ab}\, | \chi\rangle_n,$ and  $Q_w\,| \chi\rangle_n$, respectively, too.

We conclude, from the above analysis, that if the degree of a form is identified with the ghost number of a state in the quantum Hilbert space of states, then, the operation of the cohomological operators $(d,\,\delta,\,\Delta)$ is exactly like the operations of a pair of charges $(Q_b,\,Q_{ad}),\,\,(Q_d,\,Q_{ab}),\,\,Q_w\,=\,\{Q_b,\,Q_d\}\equiv\,-\,\{Q_{ad},\,Q_{ab})$ on a quantum state with the ghost number $n$. This observation can be mathematically expressed by the following mapping:
\begin{eqnarray}
(Q_b,\,Q_{ad})\rightarrow \,d, \qquad\quad (Q_d,\,Q_{ab})\rightarrow \,\delta, \qquad\quad Q_w\,=\,\{Q_b,\,Q_d \}\,\equiv -\,\{Q_{ab},\,Q_{ad}\} \rightarrow \Delta.
\end{eqnarray}
Thus, we have explained all the cohomological properties of the de Rham cohomological operators in the language of  continuous and discrete symmetry transformations of our present 2D Abelian 1-form {\it massive}
gauge theory. Whereas the continuous symmetry transformations (and corresponding generators) provide the physical realizations of the cohomological operators, it is the discrete symmetry transformations (29) that capture the properties of the Hodge duality $(*)$ operation of differential geometry. Hence, our present 2D Abelian 1-form massive gauge theory is a perfect field theoretic model for the Hodge theory.

We wrap up this section with a remark that, under the discrete symmetry transformations (29), it can
be checked that
\begin{eqnarray}
&& * \, Q_b = + \, Q_d, \qquad \quad * \, Q_d = + \, Q_b, \qquad \quad * \, Q_{ab} = + \, Q_{ad}, \nonumber\\
&& * \, Q_w = + \, Q_w, \qquad \quad * \, Q_g = - \, Q_g, \qquad \quad *\, (*\, Q_r) = + \, Q_r, 
\end{eqnarray}
where $r = b, ab, d, ad, w, g$. Thus, we note that the operation of a couple of discrete symmetry transformations
(i.e. the analogue of the Hodge duality $*$ operation of differential geometry) does not change the expression for 
the conserved charges of our present theory. However, a single operation of the Hodge duality $*$ operation changes
the charges in a suitable fashion so that the algebraic structure (33) does {\it not} change at all. The above
observations give us the clue to state that any arbitrary number of operations of the analogue of the 
Hodge duality $*$   operation (i.e. the discrete symmetry transformations (29)) does not change the algebraic structure
(33) of the conserved charges. This observation establishes the presence of a {\it perfect} duality symmetry
in our present 2D theory.

\section{Conclusions}

In our present investigation, we have mainly concentrated on the symmetries of the
modified version of 2D Proca theory and have demonstrated that the theory is endowed
with a set of $six$ continuous symmetries as well as a couple of discrete symmetry transformations.
The infinitesimal continuous symmetry transformations (and corresponding Noether conserved charges)
provide the physical realizations of the de Rham cohomological operators of differential geometry.
A couple of discrete symmetry transformations correspond to the Hodge duality ($*$) operation of
differential geometry thereby providing the physical realization  of the latter. To sum up, our
present 2D {\it massive} gauge theory provides a tractable field theoretic model for the Hodge theory
within the framework of BRST formalism.

One of the key observations in our theory is the introduction of a pseudo-scalar field ($\tilde \phi$)
[cf. (10),(11)] on the mathematical grounds of differential geometry. The presence of the discrete symmetries
in the theory fix {\it uniquely} all the signatures of the terms present in the $equivalent$ Lagrangian 
densities (10) and (11). For instance, the kinetic term of the pseudo-scalar field is negative. With the
backing from the Euler-Lagrange equation of motion [$ (\Box + m^2) \, \tilde \phi = 0$]
for this field [emerging out from the Lagrangian densities (10) and/or (11)], it is clear that such kind of 
pseudo-scalar particles could correspond to the candidates for the dark matter\footnote{Particles and fields
with negative kinetic energy have been discussed in literature in the realms
of quantum mechanics and quantum field theory (see, e.g. [30,31] and references therein).}. 
The emergence of such kinds
of fields is {\it nothing} new. This kind of kinetic term turned up, very naturally, in the context of 4D Abelian
2-form gauge theory [15-18] when the latter was proven to provide a model for the Hodge theory. Obviously,
such kind of particles have not yet been detected by the experiments in high energy physics. In our 2D massive gauge
theory, its existence is very natural on the firm ground of the existence of continuous as well as discrete symmetries in the theory.

In our present endeavor, the aesthetic of symmetries has played a key and decisive role. It is very interesting
to re-emphasize that every term carries a definite signature in our Lagrangian densities (10) and (11)
because of the presence of {\it six} continuous and {\it two} discrete symmetries in our theory. The presence of these
beautiful symmetries entail upon our massive 2D model to provide a tractable field theoretic example
 for the Hodge theory within the framework of BRST formalism (where {\it mass} and various kinds of 
continuous as well as discrete symmetries
 co-exist {\it together} in  a meaningful manner).

In our present investigation, we have focused only on the on-shell nilpotent (anti-)BRST and  (anti-)co-BRST
symmetries. It would be interesting to derive the off-shell nilpotent variety of these symmetries for our
present massive 2D Proca gauge theory within the framework of BRST formalism (see, e.g [32]). One of the geometrically intuitive methods
to obtain this kind of off-shell nilpotent and absolutely anticommuting symmetry transformations is the superfield
formalism [33-36]. In the immediate future, we plan to devote time on the derivation of these 
off-shell nilpotent and absolutely anticommuting symmetries
within the framework of ``augmented'' version of the above geometrical superfield formalism [37-42].

At this juncture, we would like to mention that, in an interesting and important piece of work [43], the four fermionic
(i.e. nilpotent) charges have been defined for the non-Abelian Proca theory in any arbitrary dimension of spacetime. A bosonic
charge, which happens to be the anticommutator of two specific fermionic charges, has also been obtained. This charge
has been shown to be connected with the appropriate Hamiltonian of the theory. However, in [43], there is no discussion about the
analogue the Hodge duality $(*)$ operation of differential geometry. Furthermore, the identifications of the fermionic
and bosonic charges with the de Rham cohomological operators of differential geometry have {\it not} been pointed out.
There is {\it no} analogy of the degree of a differential form in contrast to our case where the ghost number of a
quantum state (in the total quantum Hilbert space of states) has been identified with the degree of a form. The most
crucial difference, between our present work and that of [43], is the incorporation of a pseudo-scalar field in our
theory (i) which is the specific feature of our present 2D theory, and (ii) which is responsible for the existence
of the duality (discrete) symmetry transformations (29). To conclude, our present work, in its own right, has many 
novel features that are simple, interesting and beautiful.

Our present theory is very special because, in this theory, mass and gauge invariance co-exist together.
The 2D version of the Proca theory, modified with the inclusion of the Stueckelberg 
and a pseudo-scalar fields, is endowed with
many continuous and discrete symmetry transformations within the framework of BRST formalism. It would
be very nice endeavor to look for such kind of theories in physical four (3 + 1)-dimensions of spcetime
where the mass and various kinds of symmetries could co-exist together. Chern-Simons theory
and Jackiew-Pi model (see, e.g. [44,45]) in 3D and 
topologically massive gauge theories in 4D (with celebrated $B \wedge F$ term) are other massive guage theories. 
It would be interesting to apply Stueckelberg's formalism to such kind of theories and show them to be the
models for Hodge theory within the framework of BRST formalism. These are some of the issues that are under
investigation and our results would be reported elsewhere [46].\\

\noindent
{\bf Acknowledgements:} Discussion with R. Kumar, in the initial stages of our present
investigation, is thankfully acknowledged. TB is grateful to BHU-fellowship and DS thanks UGC,
Government of India, New Delhi, for financial support through RFSMS scheme, under which the present investigation
has been carried out.\\

\begin{center}
{\bf Appendix A: On supersymmetric type symmetry transformations}
\end{center}

\noindent
The 2D Lagrangian density (8) is endowed with the following infinitesimal and continuous 
supersymmetric (SUSY) type  transformations because we observe that the bosonic fields transform to
ferminic fields and vice-versa. These  transformations ($s, \bar s$) are 
\begin{eqnarray}
&& s \, A_\mu = -\, \varepsilon _{\mu\nu}\,\partial^\nu \,C,\qquad s \,C = 0,\qquad s \, \bar C = + \,i\,E,\nonumber\\ 
&& s \, E = \Box \,C, \qquad s \,(\partial \cdot A + m \,\phi) = 0, \qquad s \, \phi = 0,\nonumber\\
&& \bar s \, A_\mu = -\, \varepsilon _{\mu\nu}\,\partial^\nu\, \bar C,\qquad 
\bar s \,\bar C = 0,\qquad \bar s \,C = -\, i\,E,\nonumber\\  
&& \bar s\, E = \Box\, \bar C, \qquad \bar s \,(\partial \cdot A + m \,\phi) = 0, \qquad \bar s\, \phi = 0.
\end{eqnarray}
The above transformations are the {\it symmetry} transformations
because the Lagrangian density (8) transforms to the total spacetime derivatives:
\begin{eqnarray}
s \, {\cal L}_b = \partial_\mu \,\Bigl [ E\,\partial^\mu \,C + m \,\varepsilon^{\mu\nu} \,(m \,A_\nu \,C + \phi\, \partial_\nu\, C ) \Bigr ], \nonumber\\
\bar s \,{\cal L}_b = \partial_\mu \,\Bigl [ E\,\partial^\mu \,\bar C + m \,\varepsilon^{\mu\nu}\, (m \,A _\nu \,\bar C + \phi\, \partial_\nu\,\bar C ) \Bigr ].
\end{eqnarray}
 As a consequence, we observe that the action integral $ S = \int dx\, ({\cal L}_b) $ remains invariant. A few noteworthy points, at this stage, are as follows: First, the gauge-fixing term (owing its origin to the co-exterior derivative) remains invariant under $s (\bar s)$. Second, the fermionic SUSY type symmetry transformations 
$ s (\bar s)$ are nilpotent of order two $(s^2= 0, \bar s^2 = 0)$ if we take the massless (anti-)ghosts fields because $ \Box\, C = \Box \,C = 0 $. In other words, the existence of the nilpotent SUSY type symmetries entail upon the (anti-)ghosts fields to become massless. Third, the fermionic ($s^2 = \bar s^2 = 0 $) SUSY type symmetries are absolutely anticommuting (i.e. $s\, \bar s + \bar s\, s = 0 $) on the on-shell when $\Box \,C = \Box \,\bar C = 0 $ for the massless (anti-)ghosts fields. This proves the linear independence of $\bar s$ and $s$ in the {\it massless} limit.

There is a 
{\it caveat} for  all the above arguments, however. We note that there is {\it only} one mass parameter in the theory if we maintain the existence of the beautiful continuous and discrete symmetries in the theory. If mass is set equal to zero 
(i.e. $m = 0$), the whole theory reduces to a free {\it massless} gauge theory in two (1 + 1)-dimensions of spacetime which
has already been  shown to be a field theoretic model for the Hodge theory in our earlier works [25-28]. Thus, we
conclude that, to maintain the {\it non-triviality} of the theory, the massless condition {\it cannot} be imposed
on the theory. As a consequence, the SUSY type transformations (43) are 
{\it non-nilpotent} and they are {\it not} absolutely
anticmmuting in nature. Hence they cannot be identified with the (anti-)co-BRST symmetry transformations, under
which, the gauge-fixing term remains invariant [cf. (12)]. However, the transformations ($s, \bar s$) {\it do}
correspond to the {\it symmetry} transformations for the theory.

According to Noether's theorem, the existence of continuous symmetry transformations $s (\bar s)$ leads to the derivation of the conserved charges $ Q (\bar Q)$ as
\begin{eqnarray}
  Q &=& \int dx\, J^0_{(ad)} \equiv \int  dx \;[E\,\dot C - \dot E\, C], \nonumber\\
  \bar Q   &=& \int dx \,J^0_{(d)}\,\equiv  \int   dx \;[E\,\dot{\bar C}-\dot E\,\bar C] ,
\end{eqnarray}
which are derived from the conserved currents: 
\begin{eqnarray}
 J^\mu &=&  E\,\partial^\mu\, C +\varepsilon ^{\mu\nu}\,{( \partial\cdot A)}\, \partial_\nu \,C 
- m^2\,\varepsilon^{\mu\nu} \,A_\nu \,C,\nonumber\\
 \bar J^\mu   &=&  E\,\partial^\mu \,\bar C + \varepsilon^{\mu\nu}\,{(\partial\cdot A)}\,\partial_\nu\, \bar C 
- m^2\,\varepsilon^{\mu\nu}\,A_\nu\,\bar C.
\end{eqnarray}
The conservation laws $\partial_\mu \,J^\mu = \partial_\mu \bar J^\mu = 0$  can be proven by 
exploiting the following Euler-Lagrange equations of motion for the relevant fields of the theory, namely; 
\begin{eqnarray}
&&(\Box + m^2 )\,\bar C = 0, \quad(\Box + m^2)\,(\partial\cdot A) = 0, \quad
(\Box +m^2)\,\phi = 0,  \quad(\Box +m^2)\,A_\mu = 0,\nonumber\\
&& (\Box +m^2)\, E=0,\qquad \varepsilon^{\mu\nu} \,\partial_\nu\, E = \partial^\mu (\partial\cdot A) + m^2 A^\mu , \qquad(\Box + m^2)\,C = 0,
\end{eqnarray}
which are derived from the Lagrangian density (8).

We close this Appendix with the remarks that (i) to achieve the on-shell nilpotency,
we have introduced a pseudo-scalar field in the theory 
(cf. Sec. 3) on the mathematical as well as physical  grounds
 [cf. (10),(11)]. We have also seen, in addition, that the above new field is urgently
needed so as to have a discrete symmetry in the theory 
[cf. (29)] which turns out to correspond to the Hodge duality operation of differential geometry, and (ii) the symmetry transformations (43) are {\it not} exact SUSY
transformations because one of the key requirements of  a set of SUSY transformations, corresponding to
${\cal N} = 2$ supersymmetric theory, is {\it not} satisfied by the transformations (43). It can be readily
checked that the anticommutator of $(s, \bar s)$, acting on a field 
(i.e. ($s \,\bar s  + \bar s \,s) \Psi $), does {\it not} produce the spacetime translation
of the corresponding generic field $\Psi$ in our present 2D flat Minkowskian spacetime manifold. There
are many other crucial requirements of SUSY theory which are not satisfied by (43). Finally, we conclude that the transformations (43) are {\it neither} (ant-)co-BRST symmetries {\it nor} exact SUSY transformations. \\

\begin{center}
{\bf Appendix B: On a bosonic symmetry in the theory}\\
\end{center}

\noindent
The Lagrangian density (8) also respects the bosonic
($s_B^2 \neq 0$) symmetry transformation. These infinitesimal and continuous symmetry 
transformations ($s_B$), for the key individual and composite fields of the theory, are
\begin{eqnarray}
&& s_B\,A_\mu\;=\;i\, \varepsilon_{\mu\nu}\, (\Box\,A^\nu\,+\,m\,\partial^\nu\,\phi),\qquad s_B\,\phi\,=\,-i\,m\,E,
\qquad s_B\,(C,\bar C)\,=\,0,\nonumber\\ && s_B\,E\,=\,-i\,\Box\,(\partial\cdot A\,+\,m\,\phi),
\qquad \qquad s_B\,(\partial\cdot A)\,=\,-i\,\Box \,E,
\end{eqnarray}
where the transformation $ s_B\,A_\mu\;=\;i\, \varepsilon_{\mu\nu}\, (\Box\,A^\nu\,+\,m \,\partial^\nu\,\phi)$ can be written in terms of the components $ (A_0, A_1)$ of the gauge field $A_\mu$, as given below:
\begin{eqnarray}
s_B\,A_0\,=\,-i\,(\Box A_1 + m\,\partial_1\,\phi),\qquad \;\;\;
s_B A_1 = - i \,(\Box A_0 + m\, \partial_0\,\phi).
\end{eqnarray}
One of the decisive features of $s_B$ is the observation that the (anti-)ghost fields of the theory remain {\it unchanged } under this transformation (thereby rendering the ghost-part of the Lagrangian density 
to remain {\it invariant} under the  symmetry transformations ${s_B}$).

The infinitesimal and continuous transformations (48) are {\it symmetry} transformations because the Lagrangian density (8) transforms (to a total spacetime derivative) as
\begin{eqnarray}
s_B {\cal L}_b &=&\,\partial_\mu\,\Bigl[i\, m^2 (A^\mu\,E -\,m\,\varepsilon^{\mu\nu}\, A_\nu\,\phi )\,
-i\,m\,(\varepsilon^{\mu\nu}\,\phi\,\Box A_\nu\, -\,m \,\varepsilon^{\rho\sigma}\, A_\rho\,\partial^\mu\,A_\sigma)\; \nonumber\\ 
&+&\,i \,{(\partial\cdot A\,+\,m\,\phi)\,\partial^\mu E\,-\,i \,E\,\partial^\mu\, ( \partial\cdot A\,+\,m\phi ) } \Bigr].
\end{eqnarray}
As a consequence, it is clear that the action integral $S = \int\, dx\, {\cal L}_b$ would remain invariant under
the above bosonic symmetry transformations.
By exploiting the strength of the Noether's theorem, it can be seen that the conserved charge $(Q_B)$, corresponding to the symmetry transformations (48), is
 \begin{eqnarray}
Q_B = \int\,dx\,\Bigl[(\partial\cdot A + m\,\phi)\,\dot E - E\,\partial_0\, \{(\partial\cdot A + m\,\phi) \}].
\end{eqnarray} 
The above charge is the generator of the transformations (48). The validity of this statement can be proven
by exploiting the general definition of the relationship between a symmetry transformation and its generator.

\begin{center}
{\bf Appendix C: On symmetries of an alternative Lagrangian density}\\
\end{center}
We discuss here the symmetry properties of the Lagrangian density ${\cal L}_{(b_2)}$ [cf. (11) above]. Right at the 
beginning, we would like to make it clear that the dynamics remains {\it intact} as far as the Lagrangian densities
(10) and (11) are concerned. It can be seen explicitly that the Euler-Lagrange equations of motion derived
from the Lagrangian density (11) are same as the ones [cf. (17)] from the Lagrangian density (10). Thus, both
these Lagrangian densities are {\it equivalent} from the point of view of dynamics. In  what follows, we shall
demonstrate that all {\it six} continuous symmetry transformations, present
for the Lagrangian density (10),
are {\it also} respected by the Lagrangian density (11) (modulo sign factors).

It can be readily checked that, under the following on-shell [$(\Box + m^2) C = 0, (\Box + m^2) \bar C = 0$]
nilpotent ($s_{(a)b}^2 = 0$)  (anti-)BRST symmetry transformations $s_{(a)b}$
\begin{eqnarray}
&& s_{ab} \,A_\mu = \partial_\mu\, \bar C,\qquad  \,\,s_{ab}\, \bar C = 0,\qquad \quad s_{ab}\, C = i\,(\partial\cdot A - m \phi ),\nonumber\\
&& s_{ab}\,\phi = - m \,\bar C,\qquad\quad s_{ab}\, E = 0,\qquad \,\,s_{ab}\,(\partial\cdot A \,- m \,\phi)= (\Box + m^2)\,\bar C,\nonumber\\ 
&& s_{b}\, A_\mu = \partial_\mu \,C, \,\qquad \,\,s_b \,C = 0, \qquad \quad \, \,\, s_{b}\,\bar C = -i\,(\partial\cdot A - m\,\phi ), \nonumber\\
&&s_{b}\, \phi = - m\,C, \qquad \quad \,s_{b}\, E = 0, \qquad \quad s_{b}\, (\partial\cdot A - m\, \phi) = (\Box + m^2)\, C ,
\end{eqnarray}
the Lagrangian density ${\cal L}_{(b_2)}$ transforms as follows:
\begin{eqnarray}
s_{ab} {\cal L}_{(b_2)} = - \, \partial_\mu \,\bigl [ (\partial \cdot A - m \,\phi)\, \partial^\mu \bar C \bigr ],
\qquad
s_{b} {\cal L}_{(b_2)} = - \, \partial_\mu \,\bigl [ (\partial \cdot A - m \, \phi)\, \partial^\mu  C \bigr ].
\end{eqnarray}
The above (anti-)BRST symmetry transformations are also absolutely anticommuting on the on-shell. Further, under
the following  on-shell $[\Box + m^2) C = 0, (\Box + m^2) \bar C = 0$]
nilpotent ($s_{(a)d}^2 = 0$)  (anti-)co-BRST [or (anti-)dual-BRST] symmetry transformations:
\begin{eqnarray}
&& s_{ad}\, A_\mu = -\, \varepsilon _{\mu\nu}\,\partial^\nu \,C,\qquad
 \,s_{ad} \,C = 0,\qquad \,\,\,s_{ad}\, \bar C = + \,i\,(E +\, m\,\tilde\phi ),\nonumber\\ 
&& s_{ad}\, E = \Box \,C, \qquad  s_{ad}\,(\partial \cdot A - m \,\phi) = 0,
 \quad s_{ad}\, \phi = 0,\quad s_{ad}\, \tilde\phi= + m\, C,\nonumber\\ 
&& s_d\, A_\mu = -\, \varepsilon _{\mu\nu}\,\partial^\nu\, \bar C,\qquad
s_d \,\bar C = 0,\qquad\,\,\,s_d \,C = -\, i\,(E + m\,\tilde\phi ),\nonumber\\  
&& s_d\,E = \Box  \bar C, \quad\,s_d \,(\partial \cdot A - m \,\phi) = 0,
 \quad s_d\, \phi = 0 \quad s_d\, \tilde\phi= +\, m\,\bar C, 
\end{eqnarray}
the Lagrangian density (11) transforms to the total spacetime derivatives:
\begin{eqnarray}
s_{ad}\, {\cal L}_{{(b_2)}} &=& \partial_\mu \,\Bigl [ E\,\partial^\mu \,C + m \,\varepsilon^{\mu\nu} \,(m \,A_\nu \,C - \phi\, \partial_\nu\, C ) \Bigr ], \nonumber\\
s_d\,{\cal L}_{{(b_2)}} &=& \partial_\mu \,\Bigl [ E\,\partial^\mu \,\bar C + m \,\varepsilon^{\mu\nu}\, (m \,A _\nu \,\bar C - \phi\, \partial_\nu\,\bar C ) \Bigr ].
\end{eqnarray}
Thus, it is clear that the Lagrangian density (11) respects {\it all} four basic fermionic symmetry
transformations $s_{(a)b}$ and $s_{(a)d}$ that are {\it also} the symmetry transformations
for its counterpart Lagrangian density ${\cal L}_{(b_1)}$. Furthermore, we note that the
transformations $s_{(a)d}$ are also absolutely anticommuting (i.e. $s_d s_{ad} + s_{ad} s_d = 0$) on
the on-shell.

From the above {\it four} on-shell nilpotent ($s_{(a)b}^2 = 0, s_{(a)d}^2 = 0$) (anti-)BRST and (anti-)co-BRST
symmetry transformations, one can generate a {\it unique} bosonic symmetry transformation $s_\omega = \{s_b, s_d \}
= - \{s_{ad}, s_{ab} \}$ because the rest of the anticommutators turn out to be zero on the on-shell (i.e.
$\{ s_b, s_{ab} \} = 0, \{s_b, s_{ad} \} = 0, \{s_d, s_{ab} \} = 0, \{s_d, s_{ad} \} = 0$). The relevant
fields of the Lagrangian density (11) transform as follows under $s_w$:
\begin{eqnarray}
&& s_w\,A_\mu\;=\;i\,\varepsilon_{\mu\nu}\, (\Box\,A^\nu\,+\,m \,\partial^\nu\,\phi ) - i\, m\, \partial_\mu\,\tilde\phi,\qquad \quad s_w\,(C,\bar C)\,=\,0, \nonumber\\
&& s_w\,\phi\,= \,+ \, i\,m\,(E + m\,\tilde\phi ), \quad \qquad \qquad \qquad   
s_w\,\tilde\phi = - \, i\, m\, (\partial\cdot A - \, m\,\phi ),\nonumber\\
&& s_w\,E\,=\,-i\,\Box\,(\partial\cdot A - \,m\,\phi),\qquad\quad\,\,\qquad 
s_w\,(\partial\cdot A)\,=\,-i\,\Box \,(E + m\,\tilde\phi).
\end{eqnarray}
It is straightforward to check that, under the above bosonic symmetry transformations, the Lagrangian
density (11) transforms to a total spacetime derivative as:
\begin{eqnarray}
s_w {\cal L}_{(b_2)} &=& \partial_\mu  \Bigl[ i (\partial \cdot \, A - m \phi)\,\partial^\mu (E + m\tilde\phi) 
- i ( E + m\tilde\phi)\,\partial^\mu(\partial \cdot A - m \phi) \nonumber\\ &+& i \,m^2\, (A^\mu E 
+ \varepsilon^{\rho\sigma} A_\rho \,\partial^\mu A_\sigma) 
+ i m \,\varepsilon^{\mu\nu}(\phi\, \Box A_\nu + m^2 A_\nu \phi) \nonumber\\
&+& i\, m \,\{\phi\,\partial^\mu E + \tilde\phi\, \partial^\mu (\partial\cdot A)
- i\, m^2\, \tilde\phi \,\partial^\mu \phi\}  \Bigr].
\end{eqnarray}
The above expression demonstrates that the action integral $S = \int \, dx\, {\cal L}_{(b_2)}$ remains
invariant under the infinitesimal transformations $s_w$.

We close this Appendix with the remarks that (i) the operator form of the above transformations, along with the
infinitesimal ghost transformations (26), obey the algebraic structure exactly like (32). Thus, both the
Lagrangian densities (10) and (11) represent the field theoretic examples of a Hodge theory, and (ii) the
Noether conserved charges corresponding to the above symmetry transformations can be readily calculated as
we have done for the Lagrangian density (10) and they follow the same algebra as given in (33). It will
just be an academic exercise to repeat the same calculations, once again. Thus, we do not perform 
that exercise here as, we feel, there is no compelling reason for that.

\end{document}